\documentclass[11pt]{article}
\usepackage{vietnam,epsfig,amssymb,amsmath,graphicx}

\def\Journal#1#2#3#4{{#1}~{\bf #2},~#3~(#4)}
\def\PRL{Phys.~Rev.~Lett.}
\def\PRB{Phys.~Rev.~B}

\def\NAT{Nature}

\def\PHYSE{Physica~E}

\def\ttgq{\times 2e^2/h}
\def\tgq{2e^2/h}
\def\Isd{I}
\def\vsd{V_{\mathrm{sd}}}

\def\vgone{V_{\mathrm{g1}}}
\def\vgtwo{V_{\mathrm{g2}}}

\def\bpar{B_{\parallel}}
\def\He3{^3\mathrm{He}}
\def\pt7{0.7\times2e^2/h}
\def\ptseven{0.7\times2e^2/h}

\def\tauns{\tau_{n,\sigma}}
\def\kb{k_{B}}
\def\te{T_{o}}
\def\Sipart{S_{I}^{\mathrm{P}}}
\def\gavg{g_{\mathrm{avg}}}
\def\NF{\mathcal{N}}
\def\n{\rho_{n}}
\def\gx{G_\mathrm{X}}
\def\Reff{R_{\mathrm{eff}}}
\def\fs{f_{\mathrm{s}}}
\def\fd{f_{\mathrm{d}}}

\begin{document}
\vspace*{4cm}

\title{CURRENT NOISE IN QUANTUM POINT CONTACTS}
\author{L.\ DICARLO~\footnote[1]{These authors contributed equally to this work.},
YIMING\ ZHANG~\footnotemark[1], D.\ T.\ McCLURE~\footnotemark[1], D.\ J.\ REILLY, C.\ M.\ MARCUS}
\address{Department of Physics, Harvard University, Cambridge,
Massachusetts 02138, USA}
\author{L.\ N.\ PFEIFFER, K.\ W.\ WEST}
\address{Alcatel-Lucent, Murray Hill, NJ 07974, USA}
\author{M.\ P.\ HANSON, A.\ C.\ GOSSARD}
\address{Department of Materials, University of California, Santa Barbara, California 93106, USA}

\maketitle
\abstracts{
We present measurements of current noise in
quantum point contacts as a function of source-drain bias, gate
voltage, and in-plane magnetic field.  At zero bias,  Johnson
noise provides a measure of the electron temperature.
At finite bias,  shot noise at zero field exhibits an asymmetry related to the
0.7~structure in conductance. The asymmetry in noise evolves smoothly into the
symmetric signature of spin-resolved electron transmission at high field.
Comparison to a phenomenological model with  density-dependent level splitting
yields quantitative agreement. Additionally, a device-specific contribution to the finite-bias
noise, particularly visible on conductance plateaus (where shot noise vanishes), agrees
quantitatively with a model of bias-dependent electron heating.
}

The experimental discovery nearly two decades ago~\cite{QPC88a,QPC88b} of quantized conductance
in quantum point contacts (QPCs) suggested the realization of
an electron waveguide. Pioneering
measurements~\cite{Reznikov95,Kumar96,Liu98} of noise in
QPCs almost a decade later observed suppression of shot noise below the Poissonian value due to
Fermi statistics, as predicted by mesoscopic scattering theory~\cite{Lesovik89,Buttiker90th}.
Shot noise has since been increasingly recognized as an important
probe of quantum statistics and many-body effects~\cite{Blanter00,Martin05}, complementing dc transport. For example, shot-noise measurements have been exploited to directly observe quasiparticle charge in
strongly correlated systems~\cite{de-Picciotto97,Saminadayar97,Jehl00}, as well as to study interacting localized states in mesoscopic tunnel junctions~\cite{Safonov03} and cotunneling~\cite{Onac06} and dynamical channel blockade~\cite{Gustavsson06,Zhang07} in quantum dots.

Paralleling these developments, a large literature has emerged concerning the appearance of
an additional plateau-like feature in transport through a QPC at zero magnetic field, termed 0.7 structure.
Experiment~\cite{Thomas96,Kristensen00,Reilly01,Cronenwett02,Oliver02,Rokhinson06} and theory~\cite{Berggren96,Bruus01,Meir02,Matveev04,Ramsak05,Reilly05} suggest that 0.7 structure is a many-body spin effect. Its underlying microscopic origin remains an outstanding problem in mesoscopic physics.  This persistently unresolved issue is remarkable given the simplicity of the device.

In this article, we review our recent results~\cite{Point7,Techniques} on current noise in quantum point contacts---including shot-noise signatures of 0.7 structure and effects of in-plane field $\bpar$---and present new results on a device-specific contribution to noise that is well described by a model that includes bias-dependent heating in the vicinity of the QPC. Notably, we observe suppression of shot noise
relative to that predicted by theory for spin-degenerate transport~\cite{Lesovik89,Buttiker90th} near $\ptseven$ at $\bpar=0$, consistent with previous work~\cite{Avinun04,Roche04}. The suppression near $\ptseven$ evolves smoothly with increasing $\bpar$ into the signature of spin-resolved transmission. We find quantitative agreement
between noise data and a phenomenological model for a density-dependent level splitting~\cite{Reilly05},
with model parameters extracted solely from conductance.
In the final section, we investigate a device-specific contribution to the bias-dependent noise,
particularly visible on conductance plateaus (where shot noise vanishes),
which we account for with a model~\cite{Kumar96} of Wiedemann-Franz thermal conduction
in the reservoirs connecting to the QPC.

\begin{figure}[t!]
\center \label{fig1}
\psfig{figure=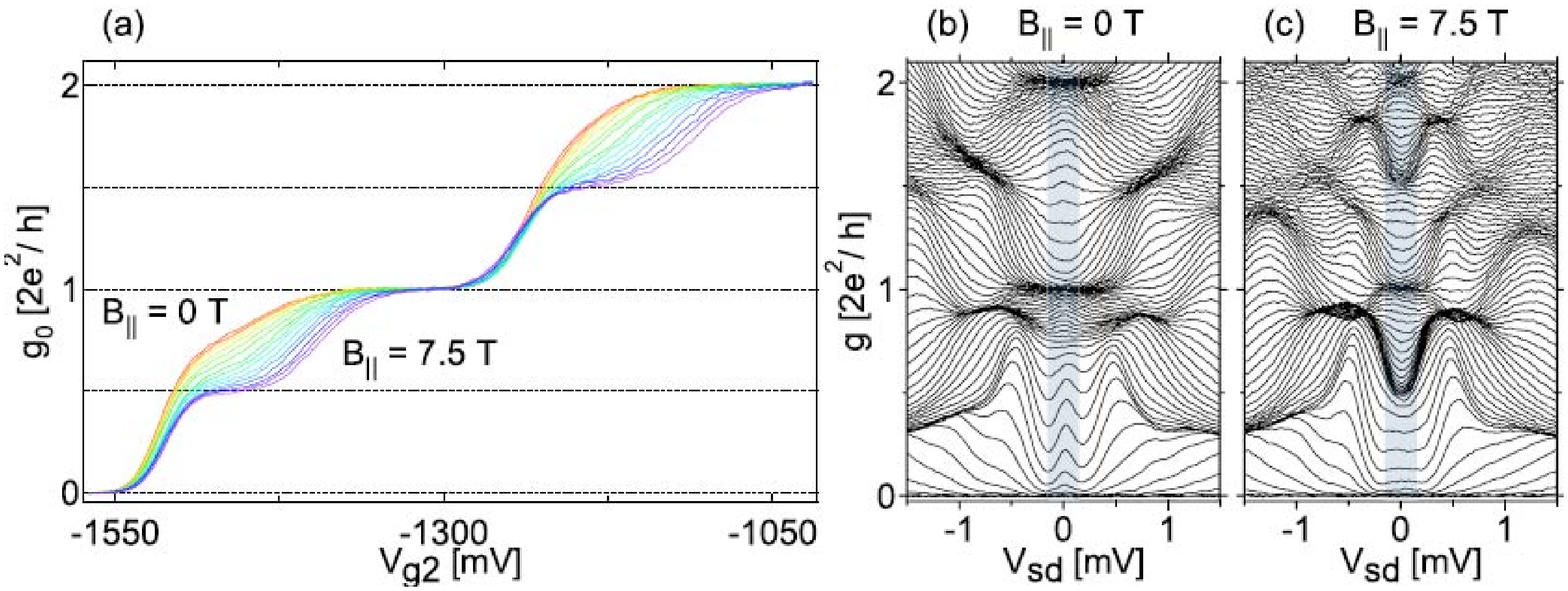,width=6.25in}
\caption{(color) (a) Linear conductance $g_0$ as a function
of $V_{\mathrm{g2}}$ ($\vgone = -\mathrm{3.2~V}$),  for $\bpar$
ranging from $0$ (red) to $7.5~\mathrm{T}$ (purple) in steps of
$0.5~\mathrm{T}$. The series resistance $R_s$ ranging from
430~$\mathrm{\Omega}$ at $\bpar=0$ to 730~$\mathrm{\Omega}$ at
$\bpar=7.5~\mathrm{T}$ has been subtracted to align the plateaus
at multiples of $\tgq$. (b,c) Nonlinear differential conductance
$g$ as a function of $\vsd$, at $\bpar= 0$ (b) and 7.5~T (c), with
$V_{\mathrm{g2}}$ intervals of 7.5 and 5~mV, respectively. Shaded
regions indicate the bias range used for the noise measurements
presented in Figs.~3(c)~and~4.
}
\end{figure}

\section*{DC transport}

Measurements are presented for two QPCs defined by split gates on
$\mathrm{GaAs}/\mathrm{Al}_{0.3}\mathrm{Ga}_{0.7}\mathrm{As}$
heterostructures grown by molecular beam epitaxy.  For QPC 1(2), the two-dimensional
electron gas [2DEG] 190(110)~nm below the heterostructure surface has  density
$1.7(2)\times 10^{11}~\mathrm{cm}^{-2}$ and mobility $5.6(0.2)\times10^6~ \mathrm{cm}^2/\mathrm{Vs}$.
Except where noted,  all data are taken at the base temperature of a $\He3$ cryostat, with electron temperature $\te$ of 290~mK.  A magnetic field of 125~mT, applied perpendicular to the plane of the 2DEG, was used to reduce bias-dependent heating~\cite{Kumar96}
(see section below).  Each QPC is first characterized at both zero and finite $\bpar$ using dc transport measurements. The differential conductance $g=dI/d\vsd$ (where $I$ is the current and $\vsd$ is the  source-drain bias) is measured by lock-in technique with an applied
$25~\mu V_{\mathrm{rms}}$ excitation at 430~Hz~\cite{Techniques}. The $\bpar$-dependent
ohmic contact and reservoir resistance $R_s$ in series with the QPC is subtracted.

Figure 1 shows conductance data for QPC 1 [see micrograph in Fig.~2(a)].
Linear-response conductance $g_0=g(\vsd\sim 0)$ as a function of gate voltage $\vgtwo$, for $\bpar=0$
to 7.5~T in steps of 0.5~T, is shown in Fig.~1(a). The QPC shows the
characteristic quantization of conductance in units of $\tgq$ at
$\bpar=0$, and the appearance of spin-resolved plateaus at multiples
of $0.5\ttgq$ at $\bpar=7.5~\mathrm{T}$. Additionally, at $\bpar=0$,
a shoulder-like 0.7 structure is evident, which evolves continuously into
the $0.5\ttgq$ spin-resolved plateau at high $\bpar$~\cite{Thomas96}.

Figures~1(b) and 1(c) show $g$ as a function of $\vsd$ for evenly spaced
$\vgtwo$ settings at $\bpar=0$ and 7.5~T, respectively.  In this representation,
linear-response plateaus in Fig.~1(a) appear as accumulated traces
around $\vsd\sim0$ at multiples of $\tgq$ for $\bpar=0$,  and at
multiples of $0.5\ttgq$ for $\bpar=7.5~\mathrm{T}$. At finite
$\vsd$, additional plateaus occur when a sub-band edge lies between the source
and drain chemical potentials~\cite{Kouwenhoven89}. The features near $0.8\ttgq$
($\vsd \sim \pm750~\mu$V) at $\bpar=0$ cannot be explained within
a single-particle picture~\cite{Patel91}. These features are
related to the 0.7 structure around $\vsd\sim0$ and resemble the
spin-resolved finite bias plateaus at $\sim0.8\ttgq$ for
$\bpar=7.5$~T~\cite{Reilly01,Cronenwett02}.

\section*{Current noise}

\begin{figure}[t]
\begin{center} \label{fig2}
\psfig{figure=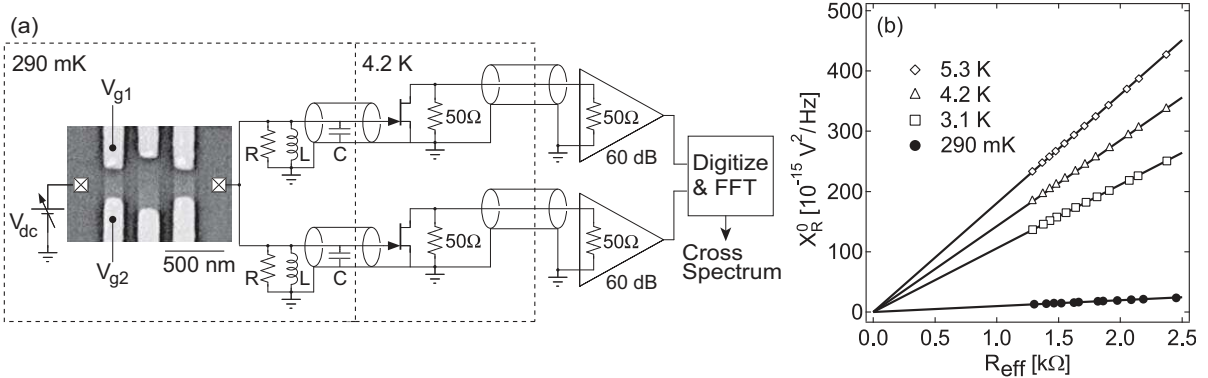,width=6.25in}
\end{center}
\caption{(a) Equivalent circuit near $2~\mathrm{MHz}$ of the system measuring QPC noise by cross-correlation on two amplification channels$^{30}$. The scanning electron micrograph shows a device of identical design to QPC 1. The QPC is formed by negative voltages $\vgone$ and $\vgtwo$ applied on two facing electrostatic gates. All other gates on the device are
grounded. (b) Calibration by Johnson noise thermometry of the electron temperature $\te$ and
the cross-correlation gain $\gx$.
$X_R^0$ as a function of $\Reff$,
at base (solid circles) and at three elevated fridge temperatures (open markers).
Solid lines are linear fits constrained to the origin. The best-fit slopes at the three elevated temperatures give $\gx=790~\mathrm{V/V}$. Combining this value of $\gx$ with the best-fit slope at base gives $\te=290~\mathrm{mK}$.}
\end{figure}

The QPC current noise is measured using a cross-correlation technique to suppress amplifier voltage noise~\cite{Techniques} [see Fig.~2(a)].
Two parallel channels amplify the voltage fluctuations across a
resistor-inductor-capacitor resonator that performs current-to-voltage conversion near the
resonant frequency of~2~MHz. Each channel consists of a transconductance stage using a
high electron mobility transistor (HEMT) cooled  to 4.2~K, followed by $50~\Omega$
amplification at room temperature. The amplified noise signals from both channels are sampled
simultaneously by a digitizer, and their cross-spectral density calculated by fast-Fourier-transform.

\subsection*{Thermal noise and calibration}

Measurement of Johnson (thermal) noise allows calibration of $\te$
and the geometric mean  $\gx$ of the voltage gain in the amplification channels, both needed to
extract bias-dependent QPC noise. The calibration procedure
illustrated in Fig.~2(b) stems from the relation  $X_{R}^{0} = \gx^{2}\cdot4\kb\te\Reff$
valid at $\vsd=0$, where $X_{R}^{0}$ is the cross-spectral density on resonance and
$\Reff$ is the effective resistance from the HEMT gates to ground.
$\Reff$ is measured  from the half-power bandwidth of the cross-spectral density~\cite{Techniques}.
At elevated refrigerator temperatures $3-5~\mathrm{K}$, where electrons are
well thermalized to a calibrated thermometer, a measurement of $X_{R}^{0}$
as a function of $\Reff$ (tuned through $\vgtwo$)
allows a calibration of $\gx$, which is found to be $790~\mathrm{V/V}$.
This value of $\gx$ is then used to determine $\te=290~\mathrm{mK}$ from the same measurement at base temperature.

\begin{figure}[b!]
\begin{center} \label{fig3}
\begin{minipage}{3.5in}
\psfig{figure=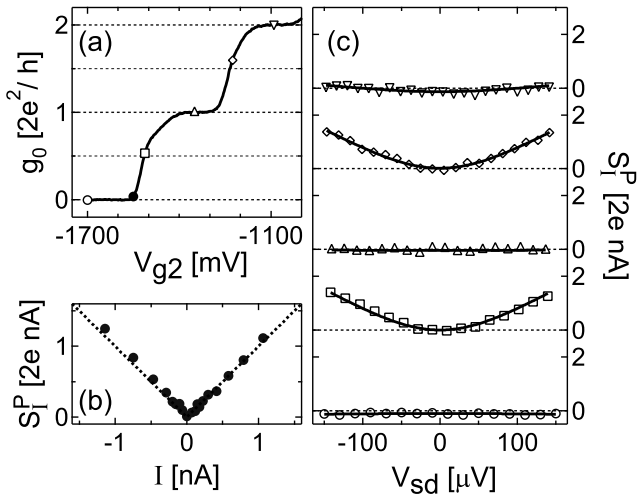,width=3.5in}
\end{minipage}\hfill
\begin{minipage}{2.5in}
\caption{
(a) Linear  conductance $g_0$ as a function
of $\vgtwo$ at $\bpar=0$. Markers indicate $\vgtwo$
settings for the noise measurements shown in (b) and (c).
(b) $\Sipart$ as a function of dc current $\Isd$ with the QPC near pinch-off. The dotted line indicates full shot noise
$\Sipart=2e|I|$.
(c) Measured $\Sipart$ as a function of
$\vsd$, for conductances near 0 (circles), 0.5 (squares), 1 (upward triangles), 1.5
(squares), and 2~$\ttgq$ (downward triangles). Solid lines are best-fits to Eq.~(2)
using $\NF$ as the only fitting parameter. In order of increasing
conductance, best-fit $\NF$ values are 0.00, 0.20, 0.00, 0.19, and
0.03.}
\end{minipage}
\end{center}
\end{figure}

\subsection*{Bias-dependent noise}

To characterize QPC noise at finite bias, we define the excess noise,
$\Sipart(\vsd)=S_{I}(\vsd)-4\kb\te g(\vsd)$, where
$S_{I}$ is the total QPC current noise spectral density.
Note that $\Sipart$ is the noise in excess of $4\kb\te g(\vsd)$
rather than $4\kb\te g(0)$ and thus differs from excess noise as
discussed in Refs. 3 and 32.
In the absence of $1/f$ and telegraph noise as well as
bias-dependent electron heating, $\Sipart$ originates from the electron partitioning at the QPC.

Experimental values for  $\Sipart$ are extracted from simultaneous measurements of $X_{R}^{0}$, $g$ and $\Reff$ using the relation
\begin{equation}
X_R^0=\gx^{2}\left(\Sipart \left(\frac{\Reff}{1+g R_s}\right)^2 + 4 \kb \te \Reff\right).
\end{equation}
With an integration time of  60~s, the resolution in
$\Sipart$ is $1.4\times10^{-29}~\mathrm{A}^2/\mathrm{Hz}$,
corresponding to full shot noise $2eI$ of $I\sim 40~\mathrm{pA}$.
$\Sipart$ as a function of dc current $I$ for QPC~1 with gates set to very low conductance ($g_0\sim 0.04\ttgq$)
[Fig.~3(b)]
exhibits full shot noise, $\Sipart=2e|I|$, demonstrating an absence of $1/f$ and
telegraph noise at the noise measurement frequency~\cite{Chen06}.

Figure~3(c) shows $\Sipart(\vsd)$ in the
$\vsd$ range $-150~\mu\mathrm{V}$ to $+150~\mu\mathrm{V}$ [shaded regions in Figs.~1(b) and 1(c)],
at $\bpar=0$ and $\vgtwo$ settings corresponding to
open markers in Fig.~3(a). Similar to when the QPC is fully pinched off,
$\Sipart$ vanishes on plateaus of linear conductance. This demonstrates
that  bias-dependent electron heating is  not significant in QPC 1.
In contrast, for $g\sim 0.5$ and $1.5\ttgq$,
$\Sipart$ grows with $|\vsd|$ and shows a transition from quadratic to linear dependence~\cite{Reznikov95,Kumar96,Liu98}. The linear dependence of $\Sipart$ on $\vsd$ at high bias further demonstrates the absence of noise due to resistance fluctuations. Solid curves superimposed on the $\Sipart(\vsd)$ data in Fig.~3(c) are best-fits to the form
\begin{equation}
\Sipart(\vsd) = 2\frac{2e^2}{h}\NF\left[e
\vsd\coth\left(\frac{e\vsd}{2\kb\te}\right)-2\kb\te\right],
\end{equation}
with the \textit{noise factor} $\NF$ as the only free fitting parameter. Note
that $\NF$ relates $\Sipart$ to $\vsd$, in contrast to the
Fano factor~\cite{Blanter00,Martin05}, which
relates $\Sipart$ to $I$. This fitting function is
motivated by mesoscopic scattering theory~\cite{Lesovik89,Buttiker90th,Blanter00,Martin05},
where transport is described by transmission coefficients
$\tauns$ ($n$ is the transverse mode index and $\mathrm{\sigma}$ denotes spin)
and partition noise originates from the partial transmission of incident electrons.
Within scattering theory, the full expression for  $\Sipart$ is
\begin{equation}
\Sipart(\vsd) =
\frac{2e^2}{h}\int\sum_{n,\sigma}\tau_{n,\sigma}(\varepsilon)(1-\tau_{n,\sigma}(\varepsilon))(\fs-\fd)^2
d\varepsilon,
\end{equation}
where $f_{\mathrm{s(d)}}$ is the Fermi function in
the source (drain) lead. Equation~(2) follows from Eq.~(3) only
for the case of constant transmission across the energy window of transport,
with  $\NF=\frac{1}{2}\sum\tauns(1-\tauns)$. Furthermore, for spin-degenerate
transmission, $\NF$ vanishes at multiples of $\tgq$ and reaches
the maximal value 0.25 at odd multiples of $0.5\ttgq$. Energy dependence of transmission
can reduce the maximal value below 0.25, as discussed below.

While Eq.~(2) is motivated by scattering theory,
the value of $\NF$ extracted from fitting to Eq.~(2) simply provides a
way to quantify $\Sipart(\vsd)$ experimentally for each $\vgtwo$.
We have chosen the bias range  $e |\vsd| \lesssim  5 \kb \te$
for fitting $\NF$ to minimize nonlinear-transport effects while
extending beyond the quadratic-to-linear
crossover in noise that occurs on the scale $e|\vsd|\sim2\kb\te$.

\begin{figure}[t!]
\center \label{fig4}
\psfig{figure=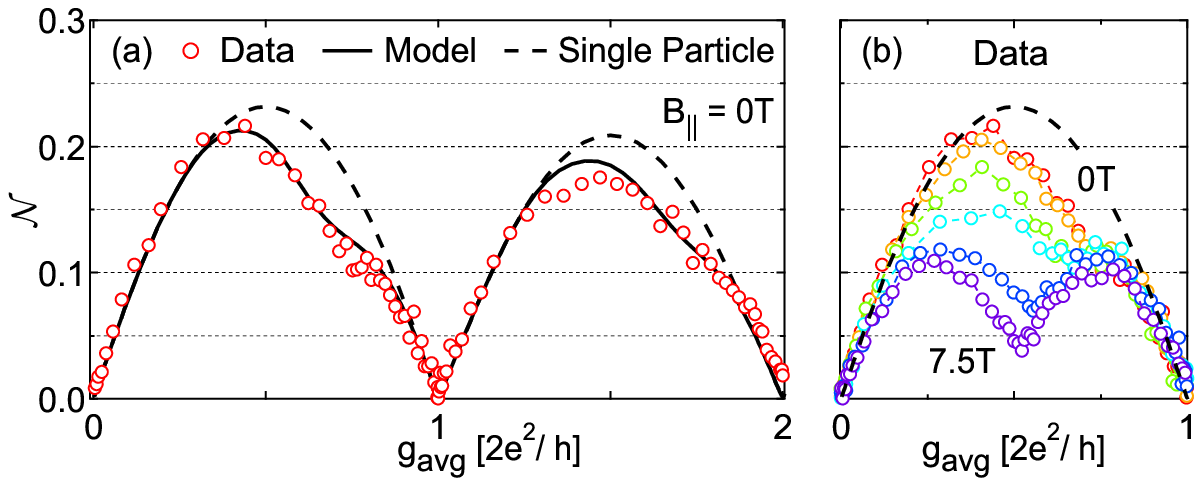,width=5.5in}
\caption{(color) (a) Experimental $\NF$
as a function of $g_{\mathrm{avg}}$ at $\bpar=0$ (red circles) along with
model curves for nonzero (solid) and zero (dashed) proportionality
of splitting, $\gamma_{n}$ (see text). (b) Experimental $\NF$ as a
function of $g_{\mathrm{avg}}$ in the range $0-1\ttgq$,
at $\bpar = $ 0~T (red), 2~T (orange), 3~T (green), 4~T (cyan), 6~T (blue),
and 7.5~T (purple). The dashed curve
shows the single-particle model ($\gamma_{n} = 0$) at zero field for comparison.
}
\end{figure}

The dependence of $\NF$ on conductance at $\bpar=0$ is shown in
Fig.~4(a), where $\NF$ is extracted from measured $\Sipart(\vsd)$ at 90
values of $\vgtwo$. The horizontal axis, $\gavg$, is the
average of the differential
conductance over the bias points where noise was measured.
 $\NF$ has the shape of a dome,
reaching a maximum near odd multiples of $0.5\ttgq$ and vanishing at
multiples of $\tgq$. The observed $\NF (\gavg)$
deviates from the spin-degenerate, energy-independent scattering theory in two
ways. First, there is a reduction in the maximum amplitude of $\NF$ below
$0.25$. Second, there is an asymmetry in $\NF$ with
respect to $0.5\ttgq$, resulting from a noise reduction near the
0.7 feature. A similar but weaker asymmetry is observed
about $1.5\ttgq$. The reduction in the maximum amplitude can be understood as resulting from an energy dependence of transmissions $\tau_{n,\sigma}$; the asymmetry is a signature of 0.7 structure, as we now discuss.

\subsection*{0.7 structure}

We investigate further the relation between the asymmetry in $\NF$ and the 0.7 structure
by measuring the dependence of $\NF(\gavg)$ on $\bpar$. As shown in Fig.~4(b),
$\NF$ evolves smoothly from a single asymmetric dome at $\bpar=0$ to a symmetric double dome at
$7.5~\mathrm{T}$. The latter is a signature of spin-resolved electron transmission.
Notably,  for $\gavg$ between 0.7 and 1 (in units of $\tgq$), $\NF$ is insensitive  to $\bpar$, in
contrast to the dependence of $\NF$ near $0.3\ttgq$.

We compare these experimental data to the shot-noise prediction of a phenomenological model~\cite{Reilly05} for the 0.7  anomaly. This model, originally motivated by dc transport data,
assumes a lifting of the twofold spin degeneracy of
mode $n$ by an energy splitting $\Delta
\varepsilon_{n,\sigma}= \sigma\cdot\n\cdot\gamma_n$ that grows linearly with
1D density $\n$ (with proportionality $\gamma_n$) within that mode. Here, $\sigma =
\pm 1$ and $\n = \sqrt{2m^*}/h \sum_{\sigma}(\sqrt{\mu_{\mathrm{s}}-\varepsilon_{n,\sigma}}+
\sqrt{\mu_{\mathrm{d}}-\varepsilon_{n,\sigma}})$, where $\mu_{\mathrm{s(d)}}$
is the source(drain) chemical potential and $m^*$ is the electron effective mass.
Parameters of the phenomenological model are extracted solely from conductance.
The lever arm converting $\vgtwo$ to energy (and hence $\n$) as well as the transverse mode spacing are extracted from transconductance $(dg/dV_{g2})$ data [Fig.~5(a)]~\cite{Patel91}.
Using an energy-dependent transmission $\tauns(\varepsilon) = 1/(1+e^{2\pi(\varepsilon_{n,\sigma}-\varepsilon)/\hbar\omega_{x,n}})$
for a saddle-point potential~\cite{Buttiker90}, the value $\omega_{x,n}$ (potential curvature parallel to the current)
is found by fitting linear conductance below $0.5\ttgq$ (below $1.5\ttgq$
for the second mode), and $\gamma_n$ is obtained by fitting
above $0.5 (1.5)\ttgq$, where (within the model)
the splitting is largest [see Fig.~5(b)]. We find $\hbar\omega_{x,1(2)}$ is $\sim500(300)~\mu\mathrm{eV}$ and $\gamma_{1(2)}\sim0.012(0.008)~e^2/4\pi\epsilon_0$ for the first
(second) mode. Note that the splitting $2\cdot \n \cdot \gamma_n$ is two orders of
magnitude smaller than the direct Coulomb energy of electrons spaced by $1/\n$.
Using these parameters, $\Sipart(\vsd)$ is calculated
using Eq.~(3), and $\NF$ is then extracted by fitting $\Sipart(\vsd)$
to  Eq.~(2).  The calculated values of $\NF(\gavg)$ at $\bpar=0$ are shown
along with the experimental data in Fig.~4(a). For comparison we include calculation results
accounting for energy-dependent transmission without splitting ($\gamma_{n}=0$).
The overall reduction of $\NF$  arises from a variation in transmission across the
$150~\mu\mathrm{V}$ bias window (comparable to
$\hbar\omega_x$), and is a single-particle effect.
On the other hand, asymmetry of $\NF$ about $0.5$ and $1.5\ttgq$ requires  nonzero $\gamma_{n}$.

\begin{figure}[t!]
\center \label{fig5}
\psfig{figure=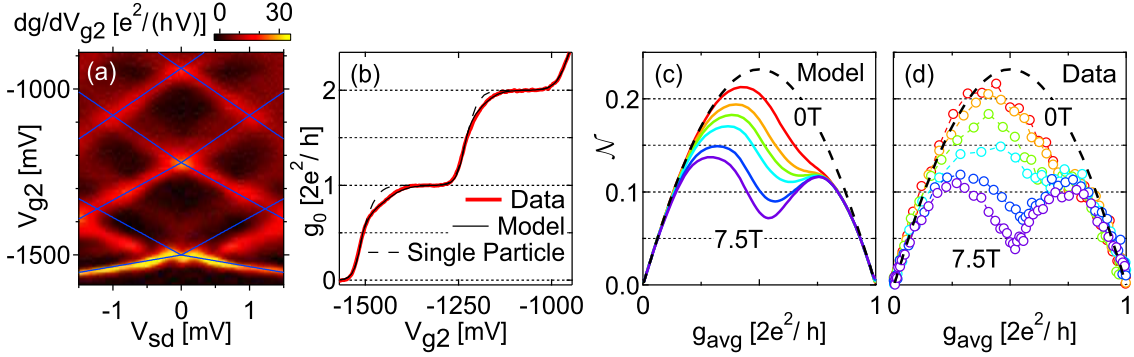,width=6.25in}
\caption{(color)
(a) Transconductance $dg/d\vgtwo$ as a
function of $\vsd$ and $\vgtwo$.
Blue lines trace the alignment of mode edges
with source and drain chemical potentials; their slope and
intersection give the conversion from $\vgtwo$ to
energy and the energy spacing between modes, respectively.
As two crossing points are observed between the first and second modes
(the model attributes this to spin-splitting in the first mode), we
take the midpoint as the crossing point for the blue lines.
(c) Measured linear conductance (red)
as a function of $\vgtwo$ at $\bpar = 0$, and linear conductance
calculated with the model (black solid) with best-fit values for $\omega_{x,n}$
and $\gamma_{n}$. Single-particle model takes $\gamma_{n} = 0$
(black dashed). (c) Model $\NF$ as a
function of $g_{\mathrm{avg}}$ in the range $0-1\ttgq$,
at $\bpar = $ 0, 2, 3, 4, 6,
and 7.5~T. (d) Same as Fig.~4(b).
}
\end{figure}

Magnetic field is included in the model by assuming a g-factor of 0.44 and adding the Zeeman splitting to the density-dependent splitting, maintaining the parameters
obtained above. Figure 5(c) shows calculated $\NF(\gavg)$ at $\bpar$ corresponding to the experimental data, reproduced in Fig.~5(d). Including the magnetic field in quadrature or as a thermally weighted mixture with the intrinsic density-dependent splitting gives essentially indistinguishable results within this model. Model and experiment show comparable evolution
of $\NF$ with  $\bpar$: the asymmetric dome for $\bpar=0$ evolves smoothly into a double dome for 7.5~T, and
for conductance $\gtrsim0.7\ttgq$, the curves for all
fields overlap closely. Some differences are observed between
data and model, particularly for  $\bpar=7.5~\mathrm{T}$.
While the experimental double dome is symmetric with respect to the
minimum at $0.5\ttgq$, the theory curve remains slightly asymmetric with a
less-pronounced minimum. We find that setting the g-factor to
$\sim0.6$ in the model reproduces the measured symmetrical
double dome as well as the minimum value of $\NF$ at $0.5\ttgq$.
This observation is consistent with reports of an enhanced
g-factor in QPCs at low density~\cite{Thomas96,Cronenwett02}.

Recent theoretical treatments of 0.7 structure have also addressed its shot-noise
signature. Modelling screening of the Coulomb interaction in the QPC, Lassl~{\it et al.}~\cite{Lassl07} qualitatively reproduce the $\bpar$-dependent $\NF$. Jaksch~{\it et al.}~\cite{Jaksch06} find a density-dependent splitting
in density-functional calculations that include exchange and correlation effects.
This theory justifies the phenomenological model and is consistent with the
observed shot-noise suppression. Using a  generalized single-impurity Anderson model motivated by density-functional calculations that suggest a quasi-bound state~\cite{Rejec06}, Golub~{\it et al.}~\cite{Golub06} find quantitative agreement with the $\bpar$-dependent $\NF$.

\subsection*{Bias-dependent electron heating}

\begin{figure}[t!]
\begin{center} \label{fig6}
\begin{minipage}{4in}
\psfig{figure=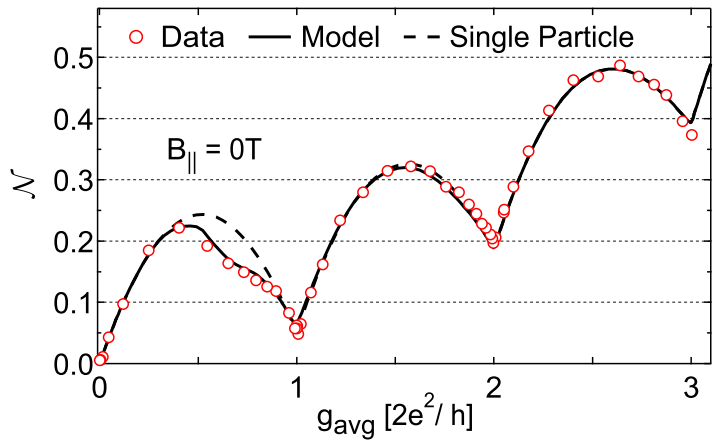,width=4in}
\end{minipage}\hfill
\begin{minipage}{2.25in}
\caption{(color)  Experimental $\NF$ as a function of $g_{\mathrm{avg}}$
at $\bpar=0$ (red circles) for QPC~2, along with
model curves for nonzero (solid) and zero (dashed) proportionality
of splitting $\gamma_{n}$. Model calculations include bias-dependent electron heating.
}
\end{minipage}
\end{center}
\end{figure}

In contrast to QPC 1, noise data in QPC 2 show evidence of bias-dependent electron heating.
Figure~6 shows $\NF(\gavg)$ at $\bpar=0$ over the first
three conductance steps, extracted from fits using Eq.~(2) to $\Sipart(\vsd)$ data over the range
$|\vsd|\leq 400~\mu\mathrm{V}$ at 50 gate voltage settings.
As in Fig.~4(a), a clear asymmetry in the noise factor is observed,
associated with enhanced noise reduction near $0.7\ttgq$. For this device, $\NF$ remains finite on conductance plateaus, showing super-linear dependence on plateau index.
This is consistent with bias-dependent thermal noise resulting from electron heating. Following
Ref.~4, we incorporate into our model the bias-dependent electron temperature
$ T_{e}(\vsd)=\sqrt{\te^2 + (24/\pi^2) (g/g_m)(1+2g/g_m)(e \vsd/ 2 \kb)^2}$, where
$g_m$ is  the parallel conductance of the reservoirs connecting to the QPC.
This expression~\cite{Kumar96} models diffusion by Wiedemann-Franz thermal conduction
of the heat flux $g\vsd^2/2$ on each side of the QPC and of Joule heating in the reservoirs, assuming ohmic contacts thermalized to the lattice at $\te$. In the absence of independent measurements of reservoir and ohmic contact resistances, we treat $1/g_m$ as a single free parameter.

Theoretical $\NF$ curves including effects of bias-dependent heating are obtained from fits to Eq.~(2) of calculated $S_I(\vsd,T_{e}(\vsd))-4\kb\te g(\vsd)$. Parameters $\omega_{x,n}=1.35,~1.13,~ 0.86~\mathrm{meV}$ and $\gamma_n=0.019,~0.008,~0~e^2/4\pi\epsilon_0$ for the first three modes (in increasing order) are extracted from conductance data. To avoid complications arising from a zero-bias anomaly~\cite{Cronenwett02} present in this device, $\gamma_{0}$ is extracted from the splitting of the first sub-band edge in the transconductance image~\cite{Reilly05}, rather than from linear conductance. Other parameters are extracted
in the same way as for QPC 1. As shown in Fig.~6, quantitative agreement with the $\NF$ data
is obtained over the three conductance steps with $1/g_m= 75~\Omega$.

In conclusion, we have presented measurements of current noise in quantum point contacts as
a function of source-drain bias, gate voltage, and in-plane magnetic field. We have observed
a shot-noise signature of the 0.7 structure at zero field, and
investigated its evolution with increasing field into the signature of spin-resolved transmission. Comparison to a phenomenological model with density-dependent level splitting yielded quantitative agreement,
and a device-specific contribution to bias-dependent noise was shown to be consistent with electron heating.

\section*{Acknowledgments}

We thank H.-A.~Engel, M.~Heiblum, L.~Levitov, and A.~Yacoby for valuable
discussions, and S.~K.~Slater, E.~Onitskansky, N.~J.~Craig, and J.~B.~Miller
for device fabrication. We acknowledge support from NSF-NSEC,
ARO/ARDA/DTO, and Harvard University.

\end{document}